\documentclass[aps,prl,twocolumn,superscriptaddress,showpacs,floatfix]{revtex4}

\usepackage{graphicx}
\usepackage{color}

\begin{document}


\title{Anomalously Hot Electrons due to Rescatter of Stimulated Raman Scattering in the Kinetic Regime}

\author{B.~J.~Winjum }
\affiliation{Department of Electrical Engineering, University of California Los Angeles, Los Angeles, CA 90095}
\author{J.~E.~Fahlen }
\affiliation{Aret\'{e} Associates, Northridge, CA 91324}
\author{F.~S.~Tsung }
\affiliation{Department of Physics and Astronomy, University of California Los Angeles, Los Angeles, CA 90095}
\author{W.~B.~Mori }
\affiliation{Department of Electrical Engineering, University of California Los Angeles, Los Angeles, CA 90095}
\affiliation{Department of Physics and Astronomy, University of California Los Angeles, Los Angeles, CA 90095}



\begin{abstract}
Using particle-in-cell simulations, we examine hot electron generation from electron plasma waves excited by stimulated Raman scattering and rescattering in the kinetic regime where the wavenumber times the Debye length ($k\lambda_D$) is $\gtrsim 0.3$ for backscatter. We find that for laser and plasma conditions of possible relevance to experiments at the National Ignition Facility (NIF), anomalously energetic electrons can be produced through the interaction of a discrete spectrum of plasma waves generated from SRS (back and forward scatter), rescatter, and the Langmuir decay of the rescatter-generated plasma waves. Electrons are bootstrapped in energy as they propagate into plasma waves with progressively higher phase velocities.
\end{abstract}

\maketitle

Stimulated Raman scattering (SRS), the decay of a light wave into a forward propagating electron plasma wave (EPW) and a forward propagating (SRFS) or backward propagating (SRBS) light wave, involves fundamental nonlinear wave-wave and wave-particle interactions. SRS continues to be studied extensively because the loss of incoming energy due to backscatter and the potential fuel preheat due to hot electrons generated by the EPW are threats to Inertial Fusion Energy (IFE) devices such as the National Ignition Facility (NIF). Recent experiments at NIF have shown electron heating up to energies above 100 keV \cite{dewald}.  A low-temperature ($T_e$ = 10 - 20 keV) part of the heated electron distribution can be attributed to SRBS, but the high-temperature part is currently unexplained.  There is speculation that these electrons are generated near the quarter-critical density by instabilities such as two-plasmon decay or SRFS \cite{michel}.  

In this article, we present a novel mechanism for generating 100 keV electrons through SRS rescatter, specifically through SRBS of SRBS, SRBS of SRFS, and the Langmuir decay instability (LDI) of the rescatter plasma waves, where LDI is the decay of an EPW into a counter-propagating EPW and an ion acoustic wave.  We further show how electrons can get progressively heated as they travel between waves of increasing phase velocities.  This mechanism allows rescatter and SRFS to heat electrons that have been initially heated by SRBS, even though the phase velocity of SRFS is too high for it to trap and heat electrons on its own.

Particle-in-cell (PIC) simulations have been used to study rescatter and multi-stage electron heating from SRS, albeit only electron heating between SRBS and SRFS.  Hinkel \textit{et al.}~\cite{hinkel} showed rescattering for NIF-relevant parameters but not the resulting hot electrons.  Other authors \cite{estabrook,mori:thesis,bertrand,langdon} have shown electron heating by SRFS, in some cases explicitly due to SRFS accelerating electrons initially heated by SRBS, but these simulations have been for more intensely-driven and/or hotter plasmas outside the current range for NIF where electron temperatures $T_e \approx$ 2-6 keV and laser intensities $I$ and wavelengths $\lambda_0$ are such that $I\lambda_0^2 \approx 10^{14}$ W $\mu$m$^2$/cm$^2$ (in laser hot spots).

\begin{figure}
\centering
\includegraphics[width=\columnwidth]{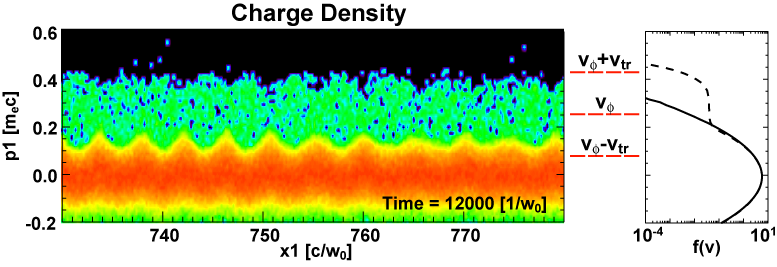}
\caption{\label{fig:dist} Electron phasespace showing trapped particles during SRBS (left) with the corresponding flattening of the distribution function (right, dashed) and initial distribution (right, solid).}
\end{figure}

The electron energies that result from trapped particle interactions in an EPW depend on its phase velocity, $v_{\phi}$, and potential amplitude, $\Phi$.  In the wave frame, the energy of a plasma electron, $(\gamma^{\prime}-1)mc^2 - e\Phi^{\prime}$, is conserved, where $\Phi^{\prime} = \gamma_{\phi}\Phi$ and $\gamma \equiv \frac{1}{\sqrt{(1-(v/c)^2)}}$.  A trapped electron with the highest energy will be at rest at the top of the potential hill in the wave frame. At the bottom of the potential such an electron has a  $\gamma^{\prime}_{max} = 1 + \frac{e\Delta\Phi^{\prime}}{mc^2}$ and $\frac{p_{max}^{\prime}}{mc} = \pm \sqrt{(1 + \frac{e\Delta\Phi^{\prime}}{mc^2})^2 - 1}$, where $\Delta\Phi$ is the potential difference between the top and bottom of the wave. $\gamma_{max}$, $p_{max}$, and $v_{max}$ can then be obtained by transforming these expressions back to the lab frame.  In the nonrelativistic limit, the maximum velocity and energy of a trapped electron are $v_{max} = v_{\phi} + \sqrt{2e(\Delta\Phi)/m}$ and $\mathcal{E}_{max} = \frac{1}{2}mv_{max}^2$.  For a sinusoidal plasma wave, $\Delta\Phi = 2\Phi_{max}$ and $v_{max} = v_{\phi} + v_{tr}$, with the trapping width $v_{tr} = 2\sqrt{e\Phi/m}$. Figure \ref{fig:dist} illustrates trapped electrons oscillating between $v_{max} = v_\phi \pm v_{tr}$ in a kinetic SRS simulation (the first simulation detailed later).

\begin{figure}
\centering
\includegraphics[width=\columnwidth]{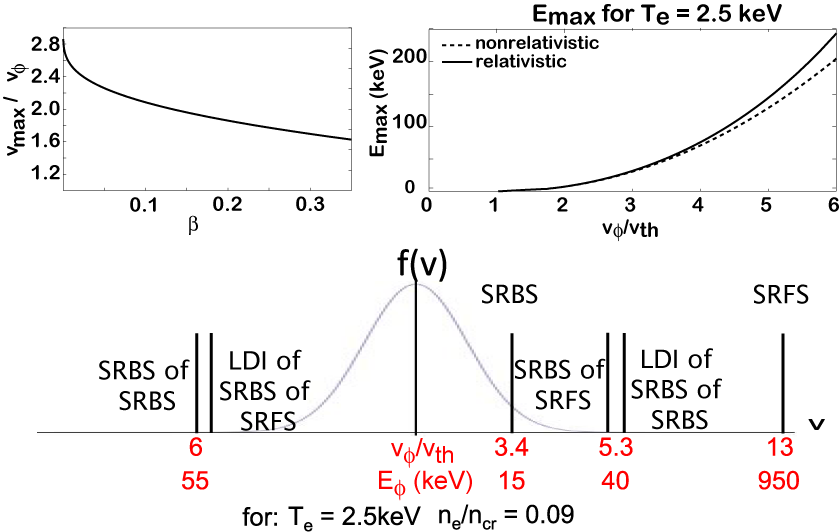}
\caption{\label{fig:ene-vph} Estimate of bound on maximum electron velocity $v_{max}$ (top-left) and kinetic energy $E_{max}$ (assuming $T_e$ = 2.5 keV, top-right).  EPW phase velocities for one set of parameters (bottom) illustrate the phase-velocity-ordering of modes, with rescatter and LDI of rescatter intermediate between SRFS and SRBS.  }
\end{figure}

One can also estimate a bound on maximum electron energy by estimating that the plasma wave is at the warm wavebreaking limit \cite{coffey}.  The wavebreaking derivation of \cite{mori:90} shows that extrema in $\Phi$,
$$
\Phi = - \overline{v} + \frac{\overline{v}^2}{2} + \frac{1}{2}\overline{v}_0^2 \frac{1}{1-\overline{v}^2} + c_0,
$$
occur for roots ($\overline{E}=0$) of:
$$
\frac{\overline{E}^2}{2} = \frac{\overline{E}_{peak}^2}{2} - \frac{\overline{v}^2}{2} + \overline{\beta} \left[\frac{1}{3}\frac{1}{(1-\overline{v})^3} - \frac{1}{2}\frac{1}{(1-\overline{v})^2} - \frac{1}{6} \right],
$$
where a waterbag distribution is assumed, $\overline{E} = eE/m\omega_pv_{\phi}$ is the normalized electric field, $\overline{v} = v/v_{\phi}$, $\overline{\beta} = 3(v_{th}/v_{\phi})^2$, and $c_0$ is an arbitrary constant.  Assuming $\overline{E}_{peak}$ is the wavebreaking amplitude, two of the roots in $v$ are approximately evenly spaced about $v_{\phi}$ and represent fluid velocities at the extrema in $\Phi$.  We can then use these two roots to calculate $\Delta\Phi$ between the extrema and substitute it into the above expressions for maximum energy.  Figure \ref{fig:ene-vph} top-left shows $v_{max}/v_{\phi}$ as a function of $\beta$, from which it is seen that for $\beta > 0.1$ ($k\lambda_D \gtrsim 0.18$) the difference between $v_{max}$ and $v_{\phi}$ is no bigger than $v_{\phi}$, i.e., $v_{max} < 2v_{\phi}$.  In addition, $\Delta\Phi = 2\Phi_{max}$ for $\beta > 0.1$ since the plasma wave is nearly sinusoidal.  Figure \ref{fig:ene-vph} top-right shows $\mathcal{E}_{max}$ assuming that $T_e$ = 2.5 keV, where the dotted line is from $\mathcal{E}_{max} = \frac{1}{2}mv_{max}^2$ and the solid line includes relativistic corrections.  In simulations, we find that $E_{peak}$ ($\Phi_{peak}$) for the SRBS wave is typically $\lesssim 2/3$ of the wavebreaking estimate, so these curves should be viewed as a limit.

The appropriate $v_{\phi}$ and $\mathcal{E}_{\phi}$ (kinetic energy for a particle at $v_{\phi}$) for the various plasma waves are shown in Figure \ref{fig:ene-vph}-bottom.  
The plots in combination show that SRBS does not longitudinally accelerate electrons to 100 keV kinetic energies.  Trapped particles with additional transverse velocity components may reach such energies, as may have occurred in L. Yin \textit{et al.}~\cite{yin:12}, but we leave this for future work. For rescatter, on the other hand, with higher $v_{\phi}$ (and lower $\beta$), 100 keV is well within range.  Rescatter EPWs, as well as their LDI decay EPWs, are potential producers of 100 keV electrons.

In this article, we simulate the scattering processes in one and two dimensions (1D and 2D) using the electromagnetic PIC code OSIRIS \cite{hemker:thesis}.  The electrons have a temperature $T_e = 2.5$ keV (3 keV) in 1D (2D) and density between $n_e = 0.09-0.10 n_{cr}$ ($k\lambda_{D} \approx 0.33$ for backward SRS); ions are either fixed or mobile with $ZT_e/T_i = 2$ and $M_i/m_e = 1836$.  The laser has normalized peak electric field $eE/mc\omega_0 = 0.0164$ which corresponds to $I_{0} = 3 \times 10^{15}$ W/cm$^{2}$ for $\lambda_{0} = 0.351 \mu$m.  In 2D the laser is focused from $I_{0} = 3 \times 10^{15}$ W/cm$^{2}$ at the simulation edge to $I_{0} = 5 \times 10^{15}$ W/cm$^{2}$ at focus with a focal spot size of 2.6 $\mu$m ($8\lambda_0$).  The laser propagates along $\hat{x}$ and is polarized in $\hat{z}$.  The simulations have absorbing boundaries for the fields and thermal-bath boundaries for the particles, with 16384 (16384 x 512) cells and 512 (256) particles per cell in 1D (2D) to simulate plasma of size 180 $\mu$m (200 x 15 $\mu$m$^2$).  The length corresponds to an $f/8$ speckle of length $8f^2\lambda_0 = 180 \mu$m.  

\begin{figure}
\centering
\includegraphics[width=\columnwidth]{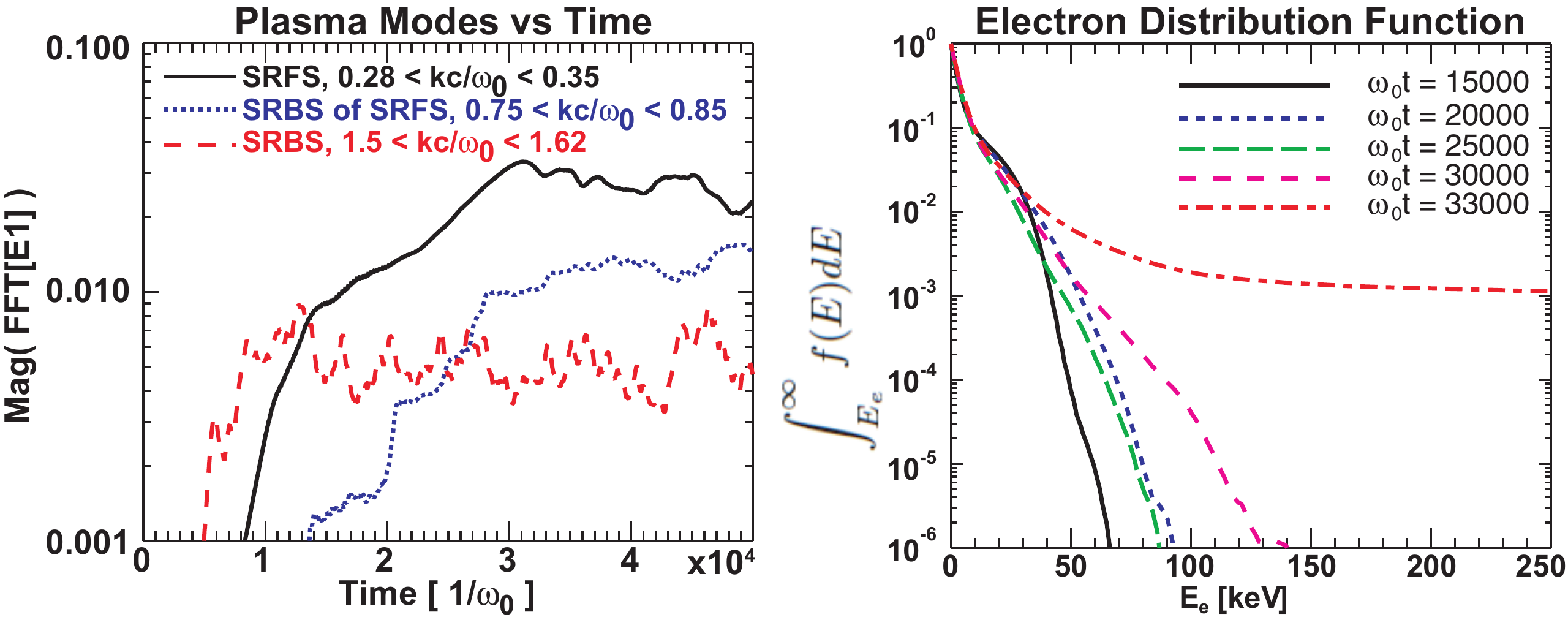}
\caption{\label{fig:dist-t-surf} Top) Temporal evolution of the SRS EPW wavenumbers. Bottom) Electron energy spectrum.}
\end{figure}

First we consider a 1D run with immobile ions and homogenous density $n_e/n_{cr} = 0.10$.  The temporal evolution of the SRS plasma wave wavenumbers can be seen in Figure \ref{fig:dist-t-surf}, along with the corresponding temporal evolution of the electron distribution function.  SRBS grows first, as it has the largest growth rate.  The growth of SRFS follows, and after SRFS has grown to sufficient amplitude for its daughter light wave to be above the SRS threshold, the SRFS saturates via SRBS of SRFS.

The hot electron tails in the distribution follow a different progression.  The daughter plasma wave phase momenta ($p_e/m_ec = \gamma v_\phi$) increase from SRBS (0.26) to SRBS of SRFS (0.56) to SRFS (2.4), with electrons at those speeds having kinetic energies of 17, 75, and 820 keV respectively.  Electron trapping by SRBS starts at $\omega_0t \approx 10000$; this process does not accelerate electrons above 70 keV.  SRFS grows to a mode amplitude larger than SRBS by $\omega_0t = 20000$, but normalized to its wavebreaking value it is smaller so it does not trap particles and has no immediate effect on the hot electron tail.  Electrons begin to be accelerated to energies above 70 keV by the rescatter that develops at $\omega_0t \approx 20000$, and by $\omega_0t \approx 33000$ electrons have been accelerated to sufficient energies by the rescatter that SRFS can interact with a significant number of electrons, trapping them and accelerating them beyond 250 keV all the way up to 1 MeV.  With heating by both rescatter and SRFS, approximately 0.1\% of the electrons get heated above 100 keV.

Though not shown, the maximum EPW amplitude in the region of SRFS activity is $eE/mv_{\phi}\omega_p \approx e\Phi/mv_{\phi}^2 \approx 0.37$ at $\omega_0t \approx 33000$.  Using $v_\phi = 0.93c$, the energy an electron needs in order to be trapped is 140 keV ($v = 0.62c$).  Since SRBS only generates electrons with energies less than approximately 60 keV, this illustrates why SRFS requires the intermediate step of rescatter.  This is consistent with Figure \ref{fig:ene-vph} where the electron energy sweeps to higher energies once rescatter heats them to 140 keV.

Rescatter can be limited by several effects.  We have performed several simulations to study its dependence on density.  Even for strong SRS, rescatter does not grow above the scattered light's quarter-critical density, which for scattered light of frequency $\omega \approx \omega_0 - \omega_p$ is $n/n_{cr} \approx 0.11$.  On the other hand, for lower densities such as $n/n_{cr} < 0.09$, the growth rates of all SRS processes decrease, likewise making rescatter less likely.  In addition, both instabilities are sensitive to density gradients and other scattering instabilities such as LDI or Brillouin scattering of the scattered light.

\begin{figure}
\centering
\includegraphics[width=\columnwidth]{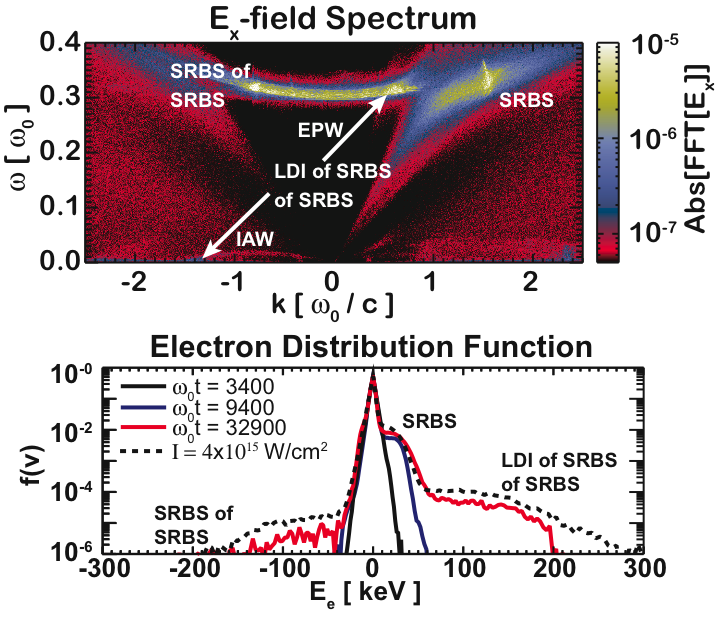}
\caption{\label{fig:grad} Top) frequency vs wavenumber of E1-field; bottom) electron distribution versus electron kinetic energy for $I_{0} = 3$ (solid) and $4$ (dashed) $\times 10^{15}$ W/cm$^{2}$, where +/- represent forward/backward traveling electrons.}
\end{figure}

While we do not see Brillouin scattering for our parameters, we do see saturation by LDI and quenching of SRFS by density gradients.  Here we turn to a 1D mobile ion simulation with a linear density gradient from $n_e/n_{cr} = 0.09$ to 0.10 over the length of the domain (180 $\mu$m).  In this simulation, SRFS does not grow.  Nevertheless, SRBS and SRBS of SRBS light still occur, as does LDI of the rescatter plasma wave.  The spectrum of plasma modes can be seen in Figure \ref{fig:grad}-top.

Figure \ref{fig:grad}-bottom shows that all of the plasma waves heat electrons, with rescatter (here SRBS of SRBS) again accelerating electrons up to energies of 100-200 keV.  Furthermore, the EPW from LDI decay of rescatter also heats electrons, and as it travels in the opposite direction as the rescatter EPW, the combined instabilities generate energetic electrons in both directions.  The LDI decay product has a slightly lower wavenumber compared to the decaying EPW, interacting with the electron distribution at slightly higher phase velocities.   The hot tail due to LDI therefore extends to higher energies than the tail due to rescatter.  Furthermore, even though the LDI EPW interacts with the electron distribution farther out in its tail, it heats more total electrons than the rescatter EPW since it also interacts with the previously formed hot tail from the original SRBS.

If we apply our earlier theoretical estimate of electron $v_{max}$ to the rescatter and LDI plasma waves, we can test this limit against the electron spectrum shown in the Figure.  The phase velocities can be ascertained from the spectrum, where the EPWs have $v_\phi \approx 5.6v_{th}$ for rescatter of SRBS and $6.4v_{th}$ for LDI of rescatter.  The electron kinetic energies corresponding to the theoretical limit (Fig. \ref{fig:ene-vph}) of $v_{max}$ are $\approx 200$ keV for SRBS and 300 keV for LDI.  These indeed bound the upper edges of the flat tails in the plotted distribution in Figure \ref{fig:grad}, even for an exactly similar case with higher laser intensity.  The low-velocity end of the hot tails correspond approximately to $\mathcal{E}_{\phi} \approx mc^2(\gamma_{\phi}-1)$ ($\approx 44$ and $61$ keV).

\begin{figure}
\centering
\includegraphics[width=\columnwidth]{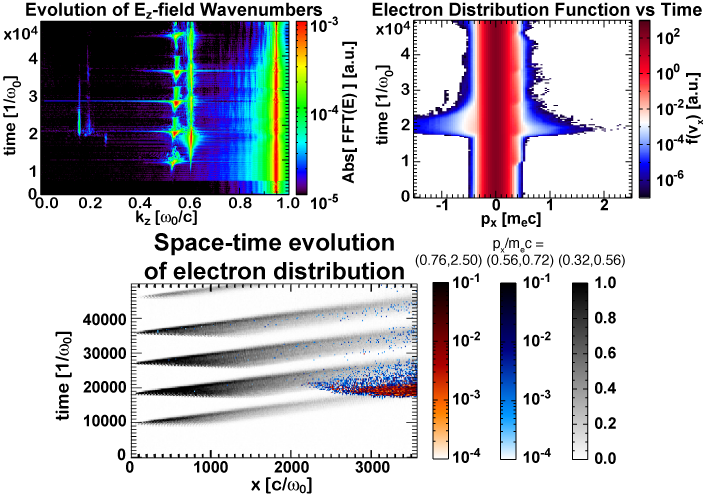}
\caption{\label{fig:2d-mobile} Top) Temporal evolution of transverse E-field wavenumbers along the simulation center (top-left) and longitudinal electron distribution spatially averaged over x and y (top-right).  Bottom) Charge density of electrons at $v_\phi$ of BSRS (grey), SRBS of SRFS (blue), and SRFS (orange).}
\end{figure}

Finally, we present results from a 2D simulation of a single speckle.  Figure \ref{fig:2d-mobile}-top-left shows the temporal evolution of wavenumbers for the $E_z$ (transverse) field along the central axis.  The bursty mode at $kc/\omega_0 \approx 0.5$ corresponds to SRBS, while the steadily growing mode at $kc/\omega_0 \approx 0.6$ that peaks at $\omega_0 t \approx 17000$ corresponds to SRFS (the anti-Stokes mode is also present at $kc/\omega_0 \approx 1.3$).  Rescatter of both light waves is present, with SRBS of SRBS at $kc/\omega_0 \approx 0.20$ and SRBS of SRFS at $kc/\omega_0 \approx 0.15$ starting at $\omega_0 t \approx 17000$.  Corresponding rescatter plasma wave modes are seen in the $E_x$ field (not shown), as well as broadband signals from modes that grow after $\omega_0 t \approx 17000$ due to LDI.

The electron distribution shown in Figure \ref{fig:2d-mobile}-top-right flattens slightly at $\omega_0 t = 10000$ due to the first burst of SRBS, followed by much more energetic tails at $\omega_0 t = 18000$ due to rescatter and LDI of rescatter.  Though not shown here, the electron phasespace reveals that the positive momentum tail is caused by SRBS of SRFS and the negative momentum tail by LDI of SRBS of SRFS.  The importance of trapped electron bootstrapping between SRBS and rescatter can be seen in the bottom plot of Figure \ref{fig:2d-mobile}, where we plot the charge density amplitude in electron phasespace as a function of space and time for three different ranges of electron momenta.  The phasespace bins $p_e/m_ec$ = (0.32, 0.56), (0.56, 0.72), and (0.76, 2.50) cover $v_\phi$ of the plasma waves due to SRBS, SRBS of SRFS, and SRFS respectively.  With SRBS growing behind the laser focus (focus at $x\omega_0/c = 1790$) and SRFS growing in front of the laser focus, the electrons trapped and accelerated by SRBS first have to cross the simulation length before interacting with the region where SRFS (and rescatter of SRFS) has grown.  After they do so (as shown in grey, the rescatter can interact with these electrons and accelerate them further.  The blue color represents electrons heated by rescatter once the electrons heated by SRBS enter the region of rescatter, while the orange color shows further acceleration by SRFS.  LDI limits SRFS for $\omega_0t > 20000$, and thereby also rescatter of SRFS.  At $\omega_0t \approx 18000$, those hot electrons with kinetic energies above 100 keV have a forward-going kinetic energy flux of approximately 3\% of the total incident laser poynting flux, while subsequent fluxes at $\omega_0t \approx 36000$ and 43000 are both $\approx 0.2$\%.

\begin{figure}
\centering
\includegraphics[width=0.45\textwidth]{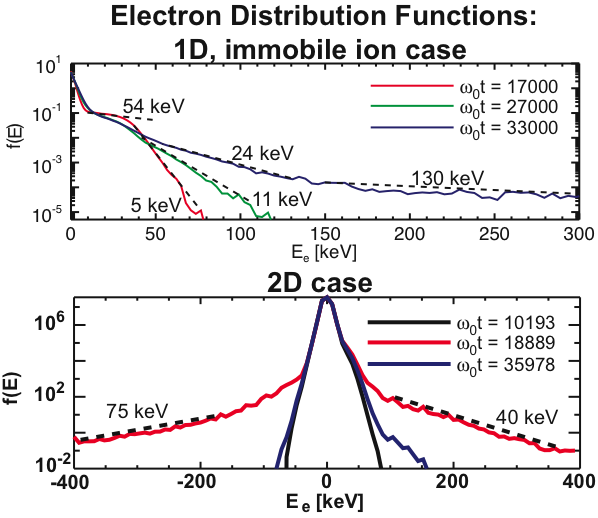}
\caption{\label{fig:temps} Energy distributions.  The 2D case includes transverse momentum, with $\pm$ referring to the sign of $p_x$.}
\end{figure}

The electron distributions in energy for the 1D immobile ion case and the 2D case are shown in Figure \ref{fig:temps}, where one can see that electrons are not accelerated to 100+ keV energies until rescatter has grown ($\omega_0t > 17000$).  Fitted lines for temperatures show that the range of electron energies, not the slope of the distribution, identifies which plasma wave (instability) is responsible for those hot electrons.

The range of energies shown in this article is consistent with reported results of hot electron measurements from NIF and shows that SRS rescatter should be considered as a source of 100 keV electrons.  While the results here are limited to single speckle physics with laser intensities at the higher end of hot spot intensities, one might reasonably assume that scattered light will be amplified to levels seen here as it travels through multiple speckles.  Yin \textit{et al.} \cite{yin:12} have shown 100 keV electrons in multi-speckle SRS, possibly resulting from electrons interacting with multiple SRBS plasma waves; however their density parameters were outside the allowable range for rescatter.  Studying rescatter in multi-speckle simulations is an area for future work. Furthermore, Yan \textit{et al.} \cite{yan:12} have shown multi-stage electron acceleration in two-plasmon-decay simulations where there occurred a broad range of plasma waves (in direction and phase velocity).  We anticipate that the bootstrap acceleration of hot electrons traveling between speckles and separate regions of SRS is worth further study, as well as the intriguing idea of whether scattered light can undergo not just SRS rescatter but also two-plasmon decay as it travels from regions above its quarter-critical density to lower densities.

This work was supported by DOE under Grant No. DE-FG52-09NA29552 and by NSF under Grant No. NSF-Phy-0904039. Simulations were carried out on the UCLA Hoffman2 cluster and NERSC's Franklin and Hopper systems.


\begin{thebibliography}{14}

\bibitem{dewald}
E. L. Dewald \textit{et al.}, Rev. Sci. Inst. {\bf 81}, 10D938 (2010).

\bibitem{michel}
P. Michel \textit{et al.}, Phys. Rev. E {\bf 83}, 046409 (2011).

\bibitem{hinkel}
D. E. Hinkel \textit{et al.}, Phys. Plasmas {\bf 11}, 1128 (2004).

\bibitem{estabrook}
K. Estabrook, W. L. Kruer, and B. F. Lasinski, Phys. Rev. Lett. {\bf 45}, 1399 (1980); K. Estabrook and W. L Kruer, Phys. Fluids {\bf 26},1892 (1983).

\bibitem{mori:thesis}
W. B. Mori, M.S. thesis, University of California, Los Angeles, 1984.

\bibitem{bertrand}
P. Bertrand \textit{et al.}, Phys. Plasmas {\bf 2}, 3115 (1995).

\bibitem{langdon}
A. B. Langdon and D. E. Hinkel, Phys. Rev. Lett. {\bf 89}, 15003 (2002).

\bibitem{coffey}
T. P. Coffey, Phys. Fluids {\bf 14}, 1402 (1971).

\bibitem{mori:90}
W. B. Mori and T. Katsouleas, Phys. Scr. {\bf T30}, 127 (1990).

\bibitem{hemker:thesis}
R. G. Hemker. Ph.D. thesis, University of California, Los Angeles, 2000;  R.A. Fonseca \textit{et al.}, Lecture Notes in Computer Science {\bf 2331}, 342 (2002); R.A. Fonseca \textit{et al.}, Plasma Physics and Controlled Fusion {\bf 50}, 124034 (2008).

\bibitem{yin:12}
L. Yin \textit{et al.}, Phys. Plasmas {\bf 19}, 056304 (2012).

\bibitem{yan:12}
R. Yan \textit{et al.}, Phys. Rev. Lett. {\bf 108}, 175002 (2012).

\end{thebibliography}
\end{document}